\documentclass[aps,prl,twocolumn,superscriptaddress,showpacs,amsmath,amssymb]{revtex4-1}
\usepackage[utf8]{inputenc}
\usepackage{amsmath,amsthm}
\usepackage{graphicx}
\usepackage[colorlinks = true,
            linkcolor = blue,
            urlcolor  = blue,
            citecolor = blue,
            anchorcolor = blue]{hyperref}
\usepackage{braket}
\usepackage{xcolor}
\newcommand{\x}{{x}}

\renewcommand{\L}{\mathbf{L}}

\newcommand{\be}{\begin{equation}}
\newcommand{\ee}{\end{equation}}
\newcommand{\bea}{\begin{eqnarray}}
\newcommand{\eea}{\end{eqnarray}}

\newcommand{\one}{\mathbb{I}}
\renewcommand{\vec}[1]{\mathbf{#1}}

\global\long\def\ga{\gamma} \global\long\def\de{\delta}
 
\global\long\def\th{\theta}
\global\long\def\th{\theta}

\global\long\def\ell#1{\theta_{#1}}
\global\long\def\bell#1{\bar{\theta}_{#1}}

 \global\long\def\ka{\kappa}
\global\long\def\si{\sigma}

\global\long\def\eps{\epsilon}
\global\long\def\al{\alpha}
\global\long\def\ga{\gamma} \global\long\def\de{\delta}

\theoremstyle{thm@}

\theoremstyle{remark}

\begin{document}

\title{Boundary driven $XY\!Z$ chain: Exact inhomogeneous triangular matrix product ansatz}
\author{Vladislav Popkov}
\affiliation{Faculty of Mathematics and Physics, University of Ljubljana, Jadranska 19, SI-1000 Ljubljana, Slovenia}
\affiliation{Bergisches Universit\"at Wuppertal, Gauss Str. 20, D-42097 Wuppertal, Germany}
\author{Xin Zhang}
\affiliation{Beijing National Laboratory for Condensed Matter Physics, Institute of Physics, Chinese Academy of Sciences, Beijing 100190, China}
\author{Toma\v z Prosen}
\affiliation{Faculty of Mathematics and Physics, University of Ljubljana, Jadranska 19, SI-1000 Ljubljana, Slovenia}

\begin{abstract}
We construct an explicit matrix product ansatz for the steady state of a boundary driven $XY\!Z$ spin-$\tfrac{1}{2}$ chain for arbitrary local polarizing channels at the chain's ends. The ansatz, where the Lax operators are written explicitly in terms of  infinite-dimensional bidiagonal (triangular) site-dependent matrices,
becomes exact either in the (Zeno) limit of infinite dissipation strength, or thermodynamic limit of infinite chain length. The solution is based on an extension of the newly discovered family of separable eigenstates of the model.
\end{abstract}
\maketitle

\textbf{ Introduction.--}
Exact and explicit solutions are indispensable for our advancement of understanding of statistical mechanics of interacting systems.
While many such exact solutions in the realm of equilibrium statistical physics are known for over a half of century~\cite{Baxter}, say the Onsager's and Baxter's solutions to two-dimensional classical statistical models at thermal equilibrium or, sometimes equivalent, Bethe-ansatz solutions for quantum models in one dimension, much less is known out of equilibrium~\cite{Gunter}.

While again, certain classical stochastic systems which are driven out of equilibrium by boundary dissipation, like simple exclusion processes, could be easily mapped to integrable quantum models in one dimension so that Bethe's ansatz can be used \cite{Mallick}, much less is understood for the corresponding boundary dissipation driven quantum lattice models \cite{JPAreview}. Only a decade ago, the first example of such a paradigm, namely the steady state density matrix of a boundary driven Lindblad master equation of the anisotropic Heisenberg ($XXZ$) spin-$\frac12$ chain has been constructed exactly~\cite{TP2011a}, paralleling analogous results for boundary driven classical stochastic lattice systems~\cite{Derrida,Blythe}.
Due to a non-compact (infinite dimensional) representation space of the fundamental algebraic objects needed in the matrix product ansatz (MPA), the solution immediately gave birth to a novel non-local, yet quasi-local conserved charge of the model~\cite{TP2011b}, resolving the long debated fundamental question on ballistic spin transport in easy-plane $XXZ$ chain at high temperature.

Using the dissipative driving in order to provide a steady-steady excitation of an otherwise conservative system in order to probe manifestly nonequilibrium physics, we will limit our discussion to jump operators which are localised at the system's boundaries. Such boundary driven paradigm can be considered as a quantum analogue of a conservative system between two thermodynamic reservoirs~\cite{Gemmer,GemmerBook}. So far, exact solutions of steady states of many-body interacting Lindblad equations were limited to particular forms of dissipative boundary driving \cite{TP2011a,TP2011b,KPS2013,Hubbarda,Hubbardb,EI2017}. Recently a new kind of MPA has been proposed, with manifestly spatially inhomogeneous matrices, in terms of which one can solve for the steady state of fully anisotropic Heisenberg ($XY\!  Z$) spin-$\frac12$ chain with arbitrarily oriented boundary polarising channels in the limit of large coupling (the so-called Zeno regime) \cite{PPZ-PRL,PPZ-PRE}.  However, the Lax matrices forming the MPA, were in general only given in terms of a numerical solution of a nonlinear recurrence in the system size, which turned unstable for long chains.

In this Letter we find an analytic solution of the aforementioned boundary driven $XY\! Z$ problem in terms of simple bidiagonal site-dependent infinite dimensional Lax operators whose elements are written explicitly in terms of Jacobi theta functions. While the solution can be formally considered to be a leading order asymptotics the Zeno regime of strong boundary coupling, it has been shown \cite{2018ZenoDynamics} that it should also apply asymptotically in the thermodynamic limit of long chains for fixed boundary coupling. Moreover, the solution provides an exact conserved charge of $XY\!Z$ model, which, unlike the Hamiltonian and the complete eight vertex transfer matrix, break the spin reversal symmetry of the model and should have applications beyond the dissipative steady state paradigm.

\textbf{Separable eigenstates of $XY\! Z$ chain.--}
We consider a chain of $N$ spins $1/2$ described by $XY\!Z$ hamiltonian $H_N$ acting over $2^N$ dimensional Hilbert space $\mathcal H_N = \mathbb C^{2^N}$
\be
H_N = \sum_{n=1}^{N-1} h_{n,n+1},\quad
h_{n,n+1} =  \sum_\alpha \sigma^\alpha_n J_\alpha \sigma^\alpha_{n+1},\quad
\ee
where $\sigma^\alpha_n$, $n\in\{1,2,\ldots N\}$, $\alpha\in \{x,y,z\}$ are Pauli operators embedded in $\mathcal H$. It turns out that the natural parametrisation of the anisotropy coupling tensor $J_\alpha$ is in terms of two complex parameters $\eta,\tau$ and Jacobi $\theta$-functions, defining shorthand notation (following \cite{WatsonBook}):
$\ell{\alpha}(u)\equiv\vartheta_\alpha(\pi u,e^{2i\pi\tau})$, $\bell{\alpha}(u)\equiv\vartheta_\alpha(\pi u,e^{i\pi\tau})$
\be
 \frac{J_x}{J} = e^{i \pi \eta} g\left({\textstyle \frac{\tau}{2}}\right),\,
 \frac{J_y}{J} = e^{i \pi \eta} g\left({\textstyle\frac{1+\tau}{2}}\right),\, 
 \frac{J_z}{J} = g\left({\textstyle \frac{1}{2}}\right),
\ee
where $g(z) \equiv \bell{1}(\eta+z)/\bell{1}(z)$.
Fixing the energy scale, say $J=1$, the remaining two independent coupling constants $J_\alpha$ are 
are uniquely parametrised – up to permutation of the axes – by taking
$\eta, {\rm i}\tau\in \mathbb R$. However, all the results of this Letter remain valid for arbitrary choice $J,\eta,\tau\in\mathbb C$ parametrizing a
general complex coupling tensor $J_\alpha$. 

Our analysis starts by the following remarkable observation. Defining a one-parameter family of spinors
\be
\ket{\psi_n}\equiv \ket{\psi(u+n\eta)} = \binom{\ell{1}(u+n\eta) }{-\ell{4}(u+n\eta)}, 
\label{eq:spinor}
\ee
where $u\in\mathbb C$ is a free parameter, we find a family of spatially inhomogeneous separable eigenstates
of $XY\! Z$ model with boundary fields:
\bea
&&(H_N-a_1 \sigma^z_1+a_N\sigma^z_N)\ket{\Psi} = E\ket{\Psi}, \label{eq:XYZa} \\ 
&&\ket{\Psi(u)} = \bigotimes_{n=1}^N \ket{\psi_n},\quad E=\sum_{n=1}^N d_n. \label{eq:sep}
\eea
The eigenvalue condition straightforwardly follows from telescoping the following divergence condition
\bea
&&h \ket{\psi_n}\otimes\ket{\psi_{n+1}} = \nonumber \\
&&(a_n \sigma^z\otimes \one_2 - a_{n+1} \one_2 \otimes \sigma^z + d_n \one_4) \ket{\psi_n}\otimes\ket{\psi_{n+1}},
\label{eq:divcond1}
\eea
where $h=\sum_\alpha J_\alpha \sigma^\alpha\otimes \sigma^\alpha$ is a $4\times 4$ hamiltonian density operator.
Consistency of (\ref{eq:divcond1}) requires that coefficients $a_n,d_n\in\mathbb C$ satisfy a set of recurrence relations which can be 
explicitly solved~\cite{SM}:
\bea
&&a_n\equiv a(u+n\eta),\; d_n = f(\eta)+f(u\!+\!n\eta)-f(u\!+\!(n\!+\!1)\eta),\nonumber\\
&&a(u) =  \frac{\bell{1}(\eta) \bell{2}(u)  }{  \bell{2}(0) \bell{1}(u) },\quad
f(u) = \frac{\bell{1}(\eta) \bell{1}'(u)  }{  \bell{1}'(0) \bell{1}(u) }\,.
\eea
Note that this fixes the magnitude of the boundary fields $a_1,a_N$, while their direction (chosen here along $z$-axis) is arbitrary so the result can be generalized to arbitrarily oriented boundary fields which not need  be collinear.

\textbf{Inhomogeneous bidiagonal Lax operators.--}
However, the choice (\ref{eq:XYZa}) serves our purpose, which is to promote Eq.~(\ref{eq:divcond1}) to a divergence relation for local Lax operators 
\cite{PPZ-PRL,PPZ-PRE}
\be
[h_{n,n+1},\mathbf{L}_n \mathbf{L}_{n+1}] = 2i(I \mathbf{L}_{n+1}-\mathbf{L}_n I).
\label{eq:divcondL}
\ee
Here $\L_n=\sum_\alpha L^\alpha_n \sigma^\alpha_n$ are the so-called Lax operators with components $L^\alpha_n\in{\rm End}(\mathcal{H}_{\rm a})$ 
as well as $I\in{\rm End}(\mathcal{H}_{\rm a})$ acting as linear operators over a suitable {\em auxiliary space}  $\mathcal{H}_{\rm a}$. 
Note that $I$ acts trivially over the physical space $\mathcal H_N$.
We first show that the solution (\ref{eq:divcond1}), together with an equivalent relation for a dual, bi-orthogonal spinor \cite{SM}
\bea
&&\bra{\psi^\perp_n} = (\theta_4(u+n\eta),\, \theta_1(u+n\eta)),\quad
 \bra{\psi^\perp_n}\otimes\bra{\psi^\perp_{n+1}}h = \nonumber \\
&& \bra{\psi^\perp_n}\otimes\bra{\psi^\perp_{n+1}}
(-a_n \sigma^z\otimes \one_2 + a_{n+1} \one_2 \otimes \sigma^z + d_n \one_4)\,,
\label{eq:divcond2}
\eea
provides a solution to (\ref{eq:divcondL}) for 1-dimensional auxiliary space $\mathcal{H}_{\rm a}=\mathbb{C}$,
\be
\L_n = \frac{1}{\ka(u+n\eta)}\ket{\psi_n}\bra{\psi_n^\perp},\quad I = 1,
\label{eq:Lax1}
\ee
where $\ka(u) = -i \theta_1(u)\theta_4(u)a(u)$. The proof follows from inserting (\ref{eq:Lax1}) into Eq.~(\ref{eq:divcondL}), while facilitating divergence conditions Eqs.~(\ref{eq:divcond1},\ref{eq:divcond2}) and a trivially verifiable identity
 $\sigma^z \ket{\psi_n}\bra{\psi_n^\perp} + \ket{\psi_n}\bra{\psi_n^\perp} \sigma^z = 2\theta_1(u+n\eta)\theta_4(u+n\eta)\one_2$.
 
 Now, we are in position to state our main result:\\
 {\bf Theorem:} {\em The operator divergence condition (\ref{eq:divcondL}) is generally solved, for any auxiliary space ${\mathcal H}_{\rm a}=\mathbb{C}^M$, with the following inhomogeneous bidiagonal ansatz
\bea
&& L^\alpha_n = \sum_{j=1}^{M} s^\alpha_{n-2(j-1)} \ket{j}\bra{j}+\sum_{j=1}^{M-1} s^\alpha_{n-2(j-1)} \ket{j}\bra{j+1}, \nonumber \\
&& I = \sum_{j=1}^{M} \ket{j}\bra{j}-\sum_{j=1}^{M-1} \ket{j}\bra{j+1}\, \label{eq:LaxM}
\eea
where $s^\alpha_n\equiv s^\alpha(u+n\eta)$ and
$s^x(u) = \frac{i}{2a(u)}\left(\frac{\ell{1}(u)}{\ell{4}(u)}-\frac{\ell{4}(u)}{\ell{1}(u)}\right)$,
$s^y(u) = -\frac{1}{2a(u)}\left(\frac{\ell{1}(u)}{\ell{4}(u)}+\frac{\ell{4}(u)}{\ell{1}(u)}\right)$,
$s^z(u) = \frac{i}{a(u)}$.}

\noindent
{\em Proof:} It is straightforward to check that (\ref{eq:LaxM}) is equivalent to  (\ref{eq:Lax1}) for $M=1$. For the general proof of (\ref{eq:LaxM}) we make the following observation:
 diagonal elements of triangular matrices (bidiagonal ones being special cases thereof) form a commutative subalgebra $\mathbb C$, hence the diagonal elements of Lax operators $\bra{j}L^\alpha_n \ket{j}$ must all have the same functional form (independent of $j$) apart from a possible shift in the variable 
$u$ (which may depend on $j$). Within the ansatz (\ref{eq:LaxM}), the matrix elements $\bra{j} L^\alpha_n \ket{j'}$ for $j'\ge j+2$ all identically vanish. Hence we only need to check the case $j'=j+1$ which is equivalent to study $2\times 2$ problem  (in auxiliary space) with 
\be
\L_n = \begin{pmatrix} 
\mathbf{u}_n & \mathbf{u}_n  \\
0 & \mathbf{v}_n \end{pmatrix},
\quad
I = \begin{pmatrix} 
1 & -1 \\
0 & 1\end{pmatrix},
\ee
where $\mathbf{u}_n = \frac{1}{\ka(u+n\eta)}\ket{\psi(u+n\eta)}\bra{\psi^\perp(u+n\eta)}$, 
$\mathbf{v}_n = \frac{1}{\ka(v+n\eta)}\ket{\psi(v+n\eta)}\bra{\psi^\perp(v+n\eta)}$. Inserting this ansatz into $\bra{1} {\rm Eq.} (\ref{eq:divcondL})\ket{2}$
and using the established identities, e.g. (\ref{eq:divcond1},\ref{eq:divcond2}), the only nontrivial condition that remains connects $u$ and $v$, i.e. $v=u-2\eta$, which proves (\ref{eq:LaxM}) for any $u,M$.
 
\textbf{Steady state of the boundary driven chain.--} 
We wish to construct the nonequilibrium steady state (NESS) density matrix $\rho$ of the Lindblad equation
\be
\frac{{\rm d}}{{\rm d}t} \rho = -i [H_{N+2},\rho] +\Gamma\,{\cal D}_{\rm l}[\rho]+\Gamma\,{\cal D}_{\rm r}[\rho]=0,
\label{eq:LME}
\ee
at large dissipation strength $\Gamma$, where ${\cal D}_{\mu}[\rho]$, $\mu\in\{{\rm l},{\rm r}\}$,  denote the dissipators at the left and right ends of the chain of $N+2$ sites, which we label by $0$ and $N+1$, respectively. They are of the form
${\cal D}_{\mu}[\rho]=2k^{}_{\mu}\rho k^\dagger_{\mu}-\{k^\dagger_{\mu}k^{}_{\mu},\rho\}$ with jump operators $k^{}_{\rm l/r}=(\mathbf{n}'_{\rm l/r} + i \mathbf{n}''_{\rm l/r})\cdot \boldsymbol{\sigma}_{0/N+1}$
targeting polarizations $\vec{n}_{\mu}=\vec{n}(\theta_{\mu},\phi_{\mu})$, where $\vec{n}(\theta,\phi)=(\sin \theta \cos \phi,\sin \theta\sin \phi,\cos \theta)$.
Here, $\vec{n}'_{\mu} = \vec{n}(\frac{\pi}{2}-\theta_{\mu},\pi+\phi_{\mu})$ and $\vec{n}''_{\mu} = \vec{n}(\frac{\pi}{2},\phi_{\mu}-\frac{\pi}{2})$, which together with $\vec{n}_{\mu}$ form an orthonormal basis of $\mathbb R^3$.
The targeted states of the dissipators are single-site pure states $\rho_{\mu}$, such that ${\cal D}_{\mu}[\rho_{\mu}]=0$.

In our previous work \cite{PPZ-PRL,PPZ-PRE,2018ZenoDynamics} we have shown that in the regime of either large $\Gamma$ or large $N$,
NESS can in the leading order be written as
$
\rho = \rho_{\rm l}\otimes \rho_N \otimes \rho_{\rm r} + {\cal O}((N\Gamma)^{-1})
$, where  
$\rho_N = \Omega \Omega^\dagger/{\rm Tr}(\Omega \Omega^\dagger)$
is completely fixed with condition
\be
\Bigl[H_N + \sum_\alpha(J_\alpha n_{\rm l}^\alpha \sigma^\alpha_1 + J_\alpha n_{\rm r}^\alpha \sigma^\alpha_N),\rho_N\Bigr]=0,
\label{eq:comm}
\ee
and the matrix product ansatz
\be
\Omega = \bra{w_{\rm l}}\L_1 \L_2 \cdots \L_N\ket{w_{\rm r}},
\ee
where $\L_n$ obey the divergence condition (\ref{eq:divcondL}). The boundary vectors $\ket{w_\mu}\in\mathcal{H}_{\rm a}$ are
fixed by projecting the commutativity conditions to the boundary sites, yielding
\bea
&&\bra{w_{\rm l}}V_{\rm l} = 0, \label{eq:BE1}\\
&&V_{\rm r} \ket{w_{\rm r}} =\varepsilon (1,-1,1,-1,\ldots)^T , 
\label{eq:BE2}\\
&&V_{\rm l}= \sum_\alpha J_\alpha n_{\rm l}^\alpha L_1^\alpha + i I, \  \ \   V_{\rm r}= \sum_\alpha J_\alpha n_{\rm r}^\alpha L_N^\alpha -i I  \nonumber
\eea
while the commutativity (\ref{eq:comm}) in the bulk follows from (\ref{eq:divcondL}).
Parameter $\varepsilon$ is arbitrary and in generic case we may fix it as $\varepsilon=1$, while a special  -- homogeneous case $\varepsilon=0$ should be treated separately.  

We do not only have a fully explicit form of the Lax operators (\ref{eq:LaxM}) but we can also solve the boundary equations explicitly (\ref{eq:BE1},\ref{eq:BE2}) and determine the free complex variable $u$.
Namely, we shall parametrize targeted boundary polarizations $n_{\rm l},n_{\rm r}$ via two complex numbers,
\be
u_\mu =x_\mu + i y_\mu,  \quad \mu\in\{{\rm l,r}\} \,, \label{eq:ulr}
\ee
as (see \cite{SM})
\begin{align}
&n_{\mu}^x = -\frac{ \bell{2}(i y_{\mu})}{\bell{3}(i y_{\mu})} \   \frac{ \bell{1}(x_{\mu})}{\bell{4}(x_{\mu})}, \nonumber\\
&n_{\mu}^y = - i \frac{ \bell{1}(i y_{\mu})}{\bell{3}(i y_{\mu})} \   \frac{ \bell{2}(x_{\mu})}{\bell{4}(x_{\mu})},  \label{eq:nvec-parametrization}\\  
&n_{\mu}^z =- \frac{ \bell{4}(i y_{\mu})}{\bell{3}(i y_{\mu})} \   \frac{ \bell{3}(x_{\mu})}{\bell{4}(x_{\mu})}\,. \nonumber
\end{align}
We can now prove, see \cite{SM} for details, that  (\ref{eq:comm}) is satisfied with the choice 
\begin{align}
& u=u_{\rm l}, \quad M=N+1\label{res:CommSolution}
\end{align}
in  (\ref{eq:LaxM}), and Eq.(\ref{eq:BE1}) is solved by 
\begin{align}
& \bra{w_{\rm l}} = \bra{1,1,0,\ldots ,0}. \label{res:wl}
\end{align}
The solution of Eq (\ref{eq:BE2}) for   $\ket{w_{\rm r}}= (r_1,r_2,\ldots, r_{N+1})^T$ for   $\eps=1$ is given by  recurrence 
 \begin{align}
&r_{k-1}=\frac{(-1)^{k} - r_{k} \,(V_{\rm r})_{k-1,k}}{(V_{\rm r})_{k-1,k-1}}, \quad  k=1, \ldots,N+1, \label{recurrence} \\
&r_{N+1}= \frac{(-1)^{N}}{(V_{\rm r})_{N+1,N+1}}.\nonumber
\end{align}
which is valid if operator $V_{\rm r}$ in (\ref{eq:BE2}) does not have zero eigenvalues, i.e.    $\prod_n (V_{\rm r})_{nn} \neq 0$.

We stress that the vector on right hand side of (\ref{eq:BE2}) should be allowed since it is essentially in the joint kernel (null space) of all $L_n^\alpha$, disregarding the last component, namely  $L_n^\alpha (1,-1,1,-1,\ldots)^T =  (0,0,\ldots,0,0,*)^T$   where $*$ denotes any nonzero element. Next action of $L_m^\beta$ creates another nonzero element 
$L_m^\beta  (0,0,\ldots,0,0,*)^T= (0,0,\ldots,0,0,*,*)^T$, and so on. The property (\ref{eq:comm}) is thus guaranteed by  
$\bra{w_{\rm l}} L_1^{\alpha_1} \ldots L_{N-1}^{\alpha_{N-1}} V_{\rm r} \ket{w_{\rm r}}= (1,1,0,0,\ldots )(0,0,*,*,\ldots ,*)^T=0 $.
In the previous study \cite{PPZ-PRL,PPZ-PRE}, on the other hand, Lax operators had a trivial joint kernel hence only $\varepsilon=0$ applied there.

\begin{figure}[tbp]
\centerline{
\includegraphics[width=0.42\textwidth]{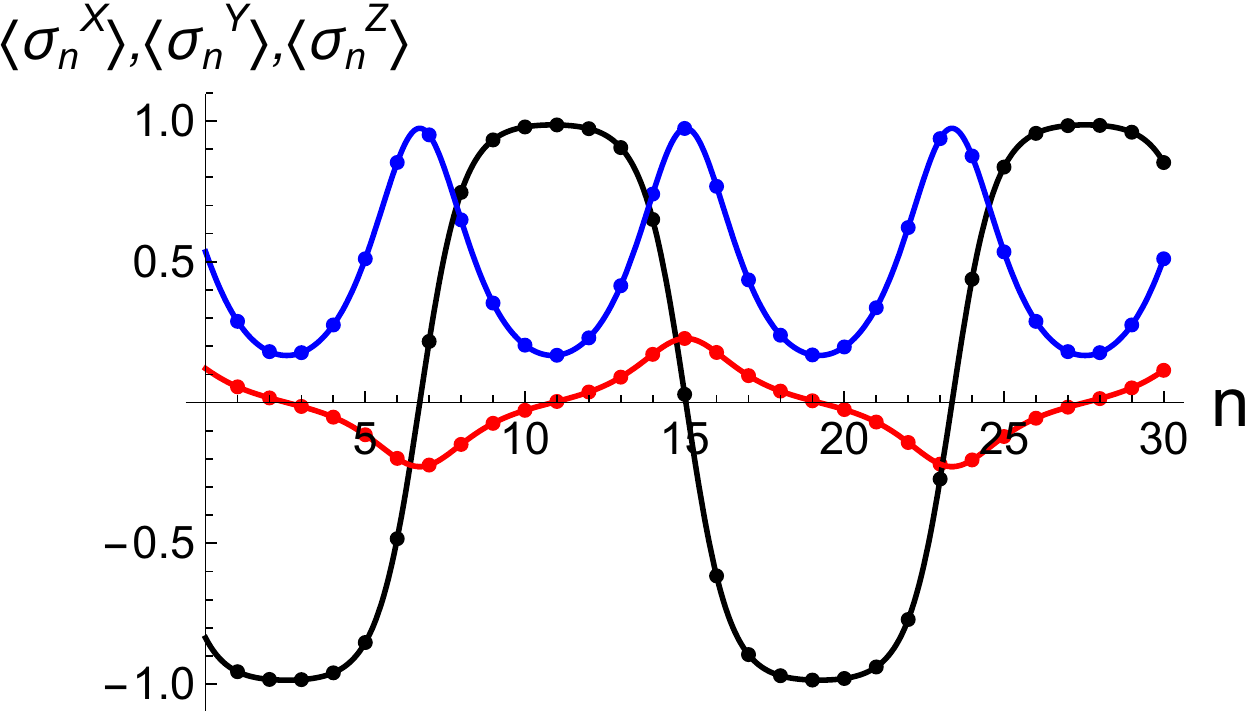}}\vspace{5mm}
\centerline
{\includegraphics[width=0.42\textwidth]{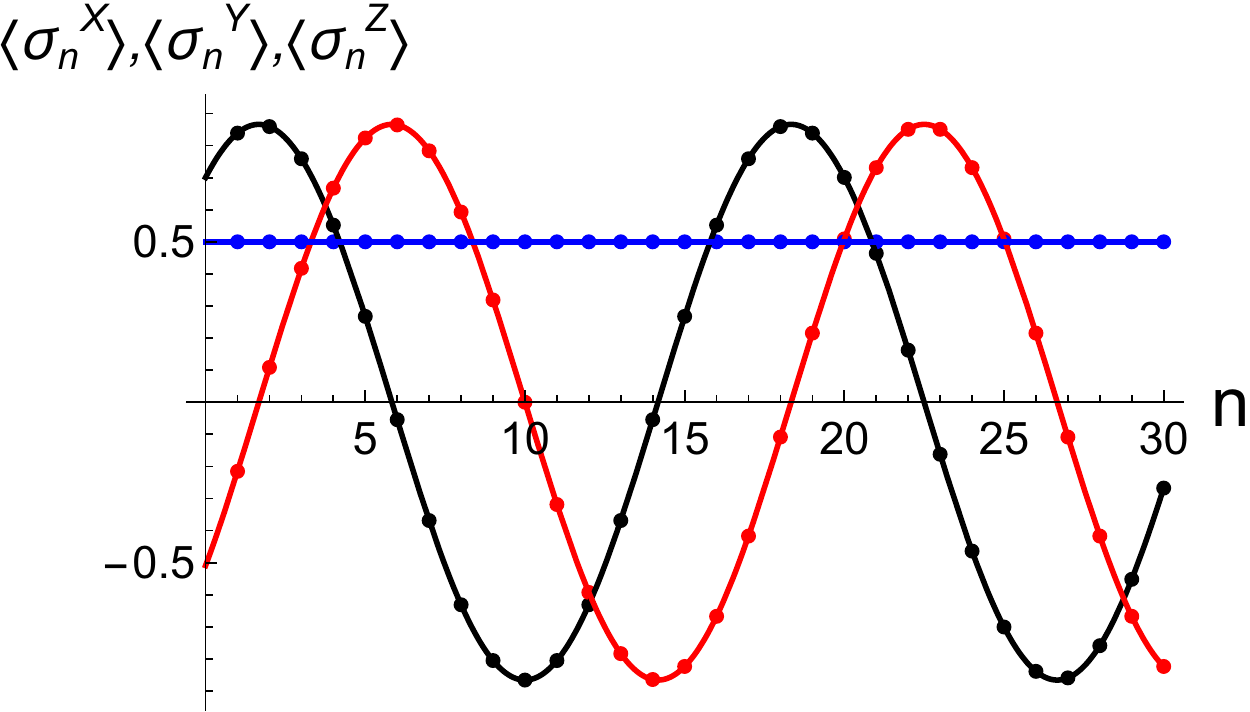}
}
\caption{Local magnetization profiles for factorized pure NESS in the 
elliptic $XY\!Z$ case (upper panel) and for  $XXZ$ case  (lower panel). $x-,y-,z-$spin projections 
are  indicated with black, red and blue points respectively. Interpolating curves for $x-,y-,z-$spin projections are given
by  Jacobi elliptic  functions (\ref{res:ell})  for $XY\!Z$ model
 and trigonometric functions for $XXZ$ model, and $n$ is site number. 
For both cases, boundary polarizations are chosen so as to render $(V_{\rm r})_{11}=0$ in Eq.~(\ref{eq:BE2}) and $N=30$, 
so that the (\ref{eq:BE2}) is solved with $\eps=0$.
   Parameters:
  $\eta=0.12,\ \tau=i/2, u=0.0477 + 0.123361 i $ (upper panel), 
  $\ga=0.12, \theta_{\rm l}=\pi/3, u=\phi_{\rm l}=-0.2 \pi$ (lower panel). }
\label{Fig-SHSelliptic}
\end{figure}

However, for a submanifold of fine tuned driving/coupling parameters the matrix $V_{\rm r}$ can be singular $\det [V_{\rm r}] = \prod_k (V_{\rm r})_{kk} = 0$~\cite{footnote}.
 Such situation  corresponds  to a nonequilibrium steady  state with large modulations of the local magnetization, see Fig.~\ref{Fig-SHSelliptic}.
In this case, the Eq.(\ref{eq:BE2}) for the right auxiliary vector $\ket{w_{\rm r}}$ must be solved  with $\eps=0$,
while the recurrence  (\ref{recurrence}) breaks down.  

The most prominent NESS of this singular type is obtained if just
the first diagonal term of $V_{\rm r}$ vanish, $(V_{\rm r})_{11}=0$, 
yielding unique solution of (\ref{eq:BE2}) with $\eps=0$:  $\ket{w_{\rm r}}= (1,0,0,\ldots, 0)^T$. The respective right 
boundary polarization $\vec{n}_{\rm r}$ is given by Eq.(\ref{eq:nvec-parametrization}) with $y_{\rm r}=y_{\rm l},\  x_{\rm r} = x_{\rm l} + (N+1)\eta $, see
\cite{SM}.
Due to upper triagonal structure of all $L_n^\al$, every 
expression of the form  $\bra{w_{\rm l}}L_1^{\al_1} L_2^{\al_2} \cdots L_N^{\al_N}\ket{w_{\rm r}}$ will contain only one nonzero term rendering the steady state site-factorized, 
\begin{align}
\rho_{N} =
 (\L_1 \L_1^\dagger) \otimes (\L_2 \L_2^\dagger) \otimes \ldots \otimes  (\L_N \L_N^\dagger) \label{SHSell}
\end{align}
 where $\L_n$ is given by (\ref{eq:Lax1}). It is easy to verify that the state is pure, and is fully characterized by the corresponding magnetization profile, 
given 
by  Jacobi elliptic  functions 
\begin{align}
&\langle  \si_n^{x}\rangle = A_\x  \ {\rm sn}(2 K_k (\eta  n + x_{\rm l} ),k), \nonumber\\
&\langle  \si_n^{y}\rangle = A_y \ {\rm cn}(2 K_k (\eta  n + x_{\rm l} ),k), \label{res:ell}\\
&\langle  \si_n^{z}\rangle = A_z \ {\rm dn}(2 K_k (\eta  n + x_{\rm l} ),k) \nonumber
\end{align}
(explicit $A_\al$ given in \cite{SM}), where $k = \left(\frac{\bell{2} (0) }{\bell{3} (0) }\right)^2$, $K_k = \frac12 \pi \ (\bell{3} (0))^2$, 
with periods $2/\eta$ ($1/\eta$) for $x,y$ ($z$) components, 
see upper panel of Fig.~\ref{Fig-SHSelliptic}. The state (\ref{SHSell}) is an elliptic counterpart of the 
spin-helix state (see lower Panel of Fig.~\ref{Fig-SHSelliptic}) appearing in models with uniaxial spin anisotropy (XXZ) \cite{2017PopkovSchutzHelix,2017PopkovPresillaJohannesJPA}. 


\textbf{ Special case of $XXZ$ chain.--}
In the partially anisotropic case $J_x/J=J_y/J=1, J_z/J=\Delta= \cos \gamma$ ($\ga$ either real or imaginary)
the divergence condition (\ref{eq:divcond1}) is satisfied with $\ket{\psi_n}= ( \cos \frac{\theta}{2} e^{-i u_n/2}, 
\sin \frac{\theta}{2} e^{i u_n/2})^T$,
$u_{n} = u+ n \gamma $ where $a_{n+1}=a_{n}= -i \sin \gamma$, $d_n= \Delta$, see \cite{SM}.
Following the same line of argument as for $XY\!Z$ case, we obtain 
explicit Lax operators in the bidiagonal form (\ref{eq:LaxM}):
\begin{align}
&s_n^x(u) \pm i s_n^y(u) = \mp\frac{  \left(\tan \frac{\theta}{2}   \right)^{\pm 1} }{\sin \ga} e^{\pm i (u + n \ga)}  
\label{sol:XXZ} \\
&s^z(u) = - \frac{1}{\sin \ga}.  \nonumber
\end{align}
The  Eq.(\ref{eq:BE1}) is satisfied with the same left boundary vector (\ref{res:wl}), 
provided that the  parameters $u,\th$ in (\ref{sol:XXZ}) relate to the spherical coordinates $\phi_{\rm l},\th_{\rm l}$
of the left boundary polarization $n_{\rm l}$ via $\th=\th_{\rm l}$, $u=\phi_{\rm l}$. The right boundary vector 
is calculated using  Eq.(\ref{eq:BE2}) with either $\eps=1$ or $\eps=0$, as discussed above.

\begin{figure}[tbp]
\centerline{
\includegraphics[width=0.42\textwidth]{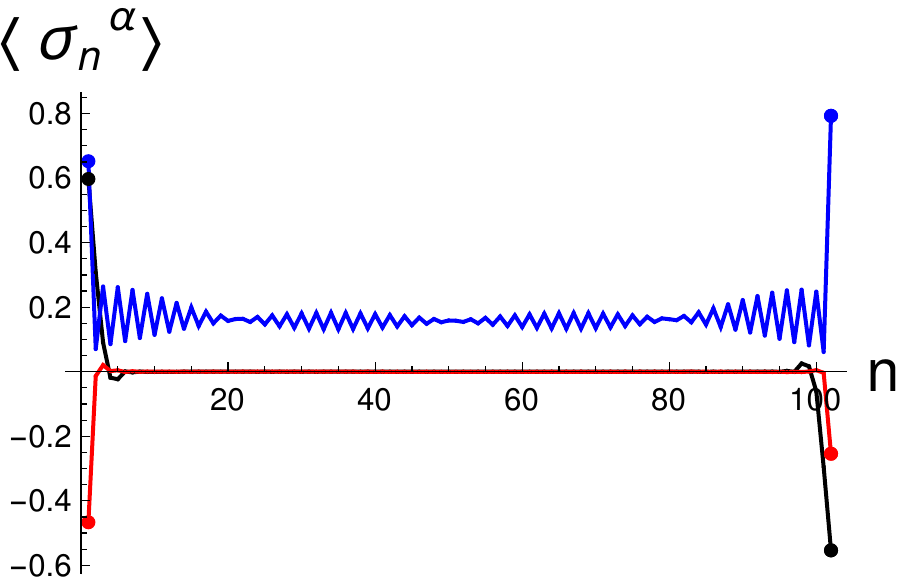}}\vspace{5mm}
\caption{ Local magnetization profiles for 
elliptic $XY\!Z$ case for a generic choice of parameters, when Eq.~(\ref{eq:BE2}) is solved with $\eps=1$, i.e. 
by recurrence (\ref{recurrence}). 
$x-,y-,z-$spin projections 
are indicated, respectively, with black, red and blue points connected by lines for clarity.  
Boundary targeted magnetizations (site numbers $n=0$, $n=N+1$) are indicated by bullet symbols at the ends.  
Parameters: $N=100$,  $\eta=0.4511,\ \tau=i/2, \ u_{\rm l}=  -0.89 +0.4 \ i, \ u_{\rm r}=  0.1 + 0.55 \ i$. 
The corresponding anisotropy tensor eigenvalues are $\{J_x,J_y,J_z\}\ = \{2.37994, 0.427449, 0.128303 \}$. 
}
\label{Fig-N100}
\end{figure}

While in special cases the NESS can be obtained fully analytically, see (\ref{res:ell}) using our MPA, 
for generic parameters simplicity and sparse structure of our MPA allows efficient numerical calculus
of arbitrary NESS observables for large chains, as exemplified in Fig.~\ref{Fig-N100}. A crucial advantage of our representation wrt to earlier results 
\cite{PPZ-PRL,PPZ-PRE} is a full control of the auxiliary vector $\ket{w_{\rm r}}$ via a recurrence (\ref{recurrence})
and explicit Lax operator expression (\ref{eq:LaxM}) for  fully anisotropic $XY\!Z$ case.

\textbf{ Discussion.--}
We have proposed an analytic method of constructing inhomogeneous MPA on the basis of local divergence condition (\ref{eq:divcond1}), by means of which we solve driven dissipative problem in the Zeno regime for a quantum spin chain, with boundary spins  kept in fixed arbitrary quantum states.
Based on our results, we identified parameters allowing to generate remarkably simple pure steady states 
with local magnetization described via Jacobi elliptic functions  (\ref{res:ell}) depicted in Fig.~\ref{Fig-SHSelliptic}.
 These states are  elliptic counterparts of spin helix states \cite{2016PRAconCarlo,2017PopkovPresillaJohannesJPA,2017PopkovSchutzHelix} discussed  recently in connection with 
cold atom experiments \cite{2020Ketterle,2021Ketterle}, from one side, and in connection with remarkable underlying algebraic structure
(phantom Bethe roots ) \cite{2021PhantomLetter, 2021PhantomLong,2021PhantomBetheAnsatz}, from the other side. 
In $XY\!Z$ model context  we can show that highly atypical quantum states of type (\ref{res:ell}) result from emergence 
of low-dimensional invariant subspaces in the spectrum of open $XY\!Z$ spin chain, under special choice of boundary fields \cite{ZKPworkinprogress}.

Our result  enable  efficient study of steady state properties of driven dissipative spin chains, reducing complexity from exponential degree $2^{2N}$ to polynomial degree $N^2$ thus allowing accessing hydrodynamic scales. From theoretical viewpoint, our construction is easily generalizable to other models
satisfying  property (\ref{eq:divcond1}), see \cite{2017PopkovSchutzHelix}, e.g. for the Izergin-Korepin model.

\let\oldaddcontentsline\addcontentsline
\renewcommand{\addcontentsline}[3]{}

\begin{acknowledgments}
We acknowledge financial support by the European Research Council through
the advanced grant No. 694544—OMNES (VP and TP), from the Deutsche
  Forschungsgemeinschaft through DFG project KL 645/20-1 (VP), and Slovenian Research Agency, Program P1-0402 (TP). 
  VP thanks the Department of Physics of Sapienza University of Rome for
  hospitality and financial support, and Carlo Presilla for discussions. 
\end{acknowledgments}


\let\addcontentsline\oldaddcontentsline 

\clearpage

\setcounter{table}{0}
\renewcommand{\thetable}{S\arabic{table}}%
\setcounter{figure}{0}
\renewcommand{\thefigure}{SM\arabic{figure}}%
\setcounter{equation}{0}
\renewcommand{\theequation}{S\arabic{equation}}%
\setcounter{page}{1}
\renewcommand{\thepage}{SM-\arabic{page}}%
\setcounter{secnumdepth}{3}
\setcounter{section}{0}
\renewcommand{\thesection}{\arabic{section}}%
\setcounter{subsection}{0}
\renewcommand{\thesubsection}{\arabic{section}.\arabic{subsection}}%
\renewcommand{\thesection}{S-\Roman{section}}

\onecolumngrid

\global\long\def\no{\nonumber}

\begin{center}\Large{
		\textit{Supplemental Material for}\\
		\medskip
		
		\textbf{Boundary driven $XY\!Z$ chain: \\
		Exact inhomogeneous triangular matrix product ansatz}\\
		
		\medskip
		by Vladislav Popkov, Xin Zhang and Toma\v z Prosen
	}
\end{center}

{
\hypersetup{linkcolor=black}
\tableofcontents
}


\bigskip

This Supplemental Material contains seven sections. In \ref{S-I} we list and prove some useful identities for elliptic theta functions. 
They are used in \ref{S-II}  to prove the general divergence condition for $XY\!Z$ Heisenberg Hamiltonian.
In \ref{S-III}  we show how the divergence condition for partially anisotropic  $XXZ$ model follows from 
the general divergence condition via a limit procedure. Sections \ref{S-IV} and  \ref{S-V} are central, as there we prove the bidiagonal Matrix Product Ansatz 
and obtain the left and the right auxiliary vectors. In \ref{S-VI} we discuss how to obtain special family of NESS which 
generalize spin-helix states to fully anisotropic $XYZ$ Hamiltonian.  In \ref{S-VII} we explicitly calculate observables 
for the simplest representative of the family, the fully factorized pure elliptic NESS.

\section{Some elliptic function identities}
\label{S-I}

In this paper, we adopt the notations of elliptic theta functions $\vartheta_{\al}(u,q)$ following Ref. \cite{WatsonBook}
\begin{align}
\begin{aligned}
&\vartheta_{1}(u,q)=2\sum_{n=0}^\infty(-1)^n q^{(n+\frac12)^2}\sin[(2n+1)u],\\
&\vartheta_{2}(u,q)=2\sum_{n=0}^\infty q^{(n+\frac12)^2}\cos[(2n+1)u],\\
&\vartheta_{3}(u,q)=1+2\sum_{n=1}^\infty q^{n^2}\cos(2nu),\\
&\vartheta_{4}(u,q)=1+2\sum_{n=1}^\infty (-1)^mq^{n^2}\cos(2nu).
\end{aligned}
\end{align}
For convenience, we use the following shorthand notations $ \ell{\al},\,\bell{\al}$ 
\begin{align}
	\ell{\al}(u) \equiv  \vartheta_{\al} (\pi u, e^{2i\pi\tau}),\quad \bell{\al}(u) \equiv   \vartheta_{\al} (\pi u, e^{i\pi\tau}),\quad {\rm Im}[\tau]>0.
\end{align}
The four types of theta functions satisfy the following identities \cite{WatsonBook}
\begin{align}
&\bell{2}(u)=\bell{1}(u+\tfrac12),\quad \bell{3}(u)=e^{i \pi(u+\frac{\tau }{4})}\bell{1}(u+\tfrac{1+\tau}{2}),\quad \bell{4}(u)=-e^{i\pi(u+\frac{\tau}{4}+\frac12)}\bell{1}(u+\tfrac{\tau}{2}),\no\\
&\ell{2}(u)=\ell{1}(u+\tfrac12),\quad \ell{3}(u)=e^{i \pi(u+\frac{\tau }{2})}\ell{1}(u+\tfrac{1+2\tau}{2}),\quad \ell{4}(u)=-e^{i\pi(u+\frac{1+\tau}{2})}\ell{1}(u+\tau).\label{FourTheta}
\end{align}
We introduce some fundamental properties of elliptic functions \cite{OffDiagonal}
\begin{align}
	&\bell{1}(-u)=\bell{1}(u),\quad \bell{1}(u+1)=-\bell{1}(u),\quad \bell{1}(u+\tau)=-e^{-2i\pi(u+\frac{\tau}{2})}\bell{1}(u),\no\\
	&\ell{1}(-u)=\ell{1}(u),\quad \ell{1}(u+1)=-\ell{1}(u),\quad \ell{1}(u+2\tau)=-e^{-2i\pi(u+\tau)}\ell{1}(u),\label{periodicity}\\
	&\th(u+x)\th(u-x)\th(v+y)\th(v-y)-\th(u+y)\th(u-y)\th(v+x)\th(v-x)\no\\
	&=\th(u+v)\th(u-v)\th(x+y)\th(x-y),\quad \th \equiv \bell{1},\,\ell{1},\label{Riemann}\\
	&\frac{\theta_1(2u)}{\theta_1(\tau)}=\frac{\bell{1}(u)\bell{2}(u)}{\bell{1}(\frac{\tau}{2})\bell{2}(\frac{\tau}{2})}=\frac{\bell{1}(u)\bell{1}(u+\frac{1}{2})}{\bell{1}(\frac{\tau}{2})\bell{1}(\frac{1+\tau}{2})},\label{Landen_1}\\
	&\frac{\bell{1}(u)}{\bell{1}(\frac{\tau}{2})}=\frac{\ell{4}(u)\ell{1}(u)}{\ell{4}(\frac{\tau}{2})\ell{1}(\frac{\tau}{2})}=e^{i\pi(u+\frac{\tau}{2})}\frac{\ell{1}(u+\tau)\ell{1}(u)}{\ell{1}^2(\frac{\tau}{2})},\label{Landen_2}
\end{align} 
Using Eq. (\ref{Riemann}) and Landen's type of transformations (\ref{Landen_1})-(\ref{Landen_2}), we get useful identities
\begin{align}
	&\frac{1}{\theta_1(u)\theta_1(u+\tau)}\left[\theta_1(u-v)\theta_1(u+v+\tau)+e^{-2i\pi v}\theta_1(u+v)\theta_1(u-v+\tau)\right]=-\frac{\theta_1(\tau)\theta_1(2v)}{\theta_1(v-\tau)\theta_1(v)}.\label{identity_1}\\
	&\bell{1}(2u)=2e^{2i\pi u}\frac{\bell{1}(u)\bell{1}(u+\frac{1}{2})\bell{1}(u+\frac{\tau}{2})\bell{1}(u+\frac{1+\tau}{2})}{\bell{1}(\frac{\tau+1}{2})\bell{1}(\frac12)\bell{1}(\frac{\tau}{2})},\label{Double_angle_1}\\
	&\ell{1}(2u)=2e^{2i\pi u}\frac{\ell{1}(u)\ell{1}(u+\frac{1}{2})\ell{1}(u+\tau)\ell{1}(u+\tau+\frac{1}{2})}{\ell{1}(\tau+\frac{1}{2})\ell{1}(\frac12)\ell{1}(\tau)}.\label{Double_angle_2}
\end{align}

Introduce the derivative of functions $\ln\ell{1}(u)$ and $\ln\bell{1}(u)$
\begin{align}
	\zeta(u)=\frac{\ell{1}'(u)}{\ell{1}(u)},\qquad \bar\zeta(u)=\frac{\bell{1}'(u)}{\bell{1}(u)}.
\end{align}
The functions $\zeta(u)$ and $\bar\zeta(u)$ possess the following properties 
\begin{align}
	&\bar\zeta(u)=-\bar\zeta(-u),\quad 
	\bar\zeta(u+1)=\bar\zeta(u),\quad \bar\zeta(u+\tau)=\bar\zeta(u)-2i\pi,\label{zeta;1}\\
	&\zeta(u)=-\zeta(-u),\quad 
	\zeta(u+1)=\zeta(u),\quad \zeta(u+2\tau)=\zeta(u)-2i\pi,\label{zeta;2}\\
	&2\zeta(2u)=\bar\zeta(u)+\bar\zeta(u+\tfrac12),\quad \bar\zeta(u)=i\pi+\zeta(u)+\zeta(u+\tau),\label{zeta;3}\\
	&2\bar\zeta(u)=2i\pi+\bar\zeta(\tfrac{u}{2})+\bar\zeta(\tfrac{u+1}{2})+\bar\zeta(\tfrac{u+\tau}{2})+\bar\zeta(\tfrac{u+\tau+1}{2}).\label{zeta;4}
\end{align}
The functions $\bell{1}(u),\,\ell{1}(u),\,\zeta(u),\,\bar\zeta(u)$ satisfy the identities 
\begin{align}
	&\frac{\bell{1}(u+\frac12)}{\bell{1}(u)}=\frac{\bell{1}(\frac12)}{\bell{1}'(0)}[\bar\zeta(\tfrac{u}{2})+\bar\zeta(\tfrac{u+1}{2})-\bar\zeta(u)],\label{zeta;sigma;1}\\
	&\frac{\bell{1}(a+b)\bell{1}(a+c)\bell{1}(b+c)}{\bell{1}(a)\bell{1}(b)\bell{1}(c)\bell{1}(a+b+c)}=\frac{1}{\bell{1}'(0)}[\bar\zeta(a)+\bar\zeta(b)+\bar\zeta(c)-\bar\zeta(a+b+c)],\label{zeta;sigma;4}\\
	&\frac{\ell{1}(a+b)\ell{1}(a+c)\ell{1}(b+c)}{\ell{1}(a)\ell{1}(b)\ell{1}(c)\ell{1}(a+b+c)}=\frac{e^{\frac{i\pi\tau}{2}}\ell{1}(\tau)\bell{1}(\frac{\tau}{2})}{\ell{1}^2(\frac{\tau}{2})\bell{1}'(0)}[\zeta(a)+\zeta(b)+\zeta(c)-\zeta(a+b+c)]\no\\
	&\overset{(\ref{Landen_2})}{=}\frac{e^{-i\pi d}\ell{1}(\tau)\bell{1}(d)}{\ell{1}(d)\ell{1}(d+\tau)\bell{1}'(0)}[\zeta(a)+\zeta(b)+\zeta(c)-\zeta(a+b+c)].\label{zeta;sigma;5}
\end{align}
for some arbitrary constants $a,b,c,d$.
Some other useful relations are:
\begin{align}
&2\ell{1} (x+y) \ell{1} (x-y) = \bell{4} (x) \bell{3} (y) -  \bell{4} (y) \bell{3} (x) \label{Derkachev1}\\
&2\ell{4} (x+y) \ell{4} (x-y) = \bell{4} (x) \bell{3} (y) +  \bell{4} (y) \bell{3} (x) \label{Derkachev2} \\
&2\ell{4} (x+y) \ell{1} (x-y) = \bell{1} (x) \bell{2} (y) -  \bell{1} (y) \bell{2} (x) \label{Derkachev3} 
\end{align}

\textbf{Proof of Eq. (\ref{zeta;sigma;1})}:
Define auxiliary functions 
\begin{align}
	c_1(u)=\frac{\bell{1}(u+\frac12)}{\bell{1}(u)},\quad c_2(u)=\frac{\bell{1}(\frac12)}{\bell{1}'(0)}\left[\bar\zeta(\tfrac{u}{2})+\bar\zeta(\tfrac{u+1}{2})-\bar\zeta(u)\right].
\end{align}
These two functions possess identical properties, as follows 
\begin{align}
	\begin{aligned}
	&\hbox{double periodicity:}\quad c_1(u+1)=c_1(u),\quad c_1(u+\tau)=-c_1(\tau),\\ 
	&\hspace{3.2cm} c_2(u+1)=c_2(u),\quad c_2(u+\tau)\overset{ (\ref{zeta;4})}{=}-c_2(\tau),\\
	&\hbox{zeros:}\quad c_1(\tfrac12)=0,\quad c_2(\tfrac12)\overset{(\ref{zeta;1})}{=}0,\\
	&\hbox{poles:}\quad c_1(0)\to\infty,\quad c_2(0)\to \infty,\\
	&\hbox{derivative at certain point:}\quad\lim_{u\to 0}\bell{1}(u) c_1(u)=\lim_{u\to 0}\bell{1}(u)c_2(u)=\bell{1}(\tfrac12).
	\end{aligned}
\end{align}
These properties imply $c_1(u)\equiv c_2(u)$. Following this method, one can prove all the other identities.  

\section{Proof of divergence condition for $XY\!Z$ model (\ref{eq:divcond1})}
\label{S-II}

The anisotropy exchange constants $J_x,J_y,J_z$ in $XY\!Z$ model are parameterized as 
\begin{align}
	J_x= e^{i \pi \eta}  \frac{\bell{1}(\eta+ \frac{\tau}{2})}  {    \bell{1}( \frac{\tau}{2})},\quad  J_y= e^{i \pi \eta}  \frac{\bell{1}(\eta+ \frac{1+\tau}{2})}  {    \bell{1}( \frac{1+\tau}{2})},\quad J_z=\frac{\bell{1}(\eta+ \frac{1}{2})}  {    \bell{1}( \frac{1}{2})},\label{Def:anisotropy}
\end{align}
where $\eta$ is generic complex number.
Introduce two parameters $J_\pm$ as 
\begin{align}
	J_\pm=J_x\pm J_y.
\end{align}

Existence of  factorized eigenstates
$\ket{\Psi} = \ket{\psi_1} \otimes \ket{\psi_2} \otimes \ldots \otimes \ket{\psi_N} $ 
in the spectrum of the $XY\!Z$ transfer matrix $T$ (8-vertex model) under special conditions (on special manifolds) has been established long ago \cite{takhtajan,Fan1996}.
Since the  transfer matrix $T$ and the $XY\!Z$ quantum spin-$\frac12$ Hamiltonian commute, $[T,\,H]=0$,
$T \Psi = \Lambda \Psi$ necessarily entails $H \Psi = E_0 \Psi$ (if the eigenvalue $E_0$ is nondegenerate).
Further, since the Hamiltonian $H$ is a sum of local terms $h_{n,n+1}$, the eigenvalue condition for a factorized state
necessarily implies a local divergence condition, which can be written as 
\begin{align}
& h \ket{\psi_n}\otimes\ket{\psi_{n+1}} = 
(a_n \sigma^z\otimes \one_2 - a_{n+1} \one_2 \otimes \sigma^z + d_n \one_4) \ket{\psi_n}\otimes\ket{\psi_{n+1}},
\label{eq:div}
\end{align}
where $h=\sum_\alpha J_\alpha \sigma^\alpha\otimes \sigma^\alpha$ is a $4\times 4$ hamiltonian density operator.
Indeed, under the condition $a_1\equiv a_{N+1}$, in a periodic system $H = \sum_{n=1}^N h_{n,n+1}$  the above entails  
\begin{align}
& H \Psi = E_0 \Psi, \quad E_0= \sum_{n=1}^N d_n.
 \end{align}

Here our aim is to prove (\ref{eq:div}) and establish compact form for the coefficients $a_n,a_{n+1},d_n$.  The Eq. (\ref{eq:div}) contains 4 scalar equations and only 3 unknowns
so there must be a consistency condition. The  consistency condition has the form
\begin{align} 
&(\bra{\psi_n^\perp} \otimes \bra{\psi_{n+1}^\perp})  h (\ket{\psi_n}  \otimes \ket{\psi_{n+1}})=0 \label{eq:consistency}\\
&\braket{\psi_n^\perp|\psi_n}=0. \nonumber
\end{align}

We make the following ansatz: 
\begin{align}
& \ket{\psi_n} = \binom{\ell{1}(u) }{-\ell{4}(u)}\equiv \ket{\psi(u)}
, \quad \ket{\psi_{n+1}} \equiv \ket{\psi(u+\eta)} \label{eq:QubitAnsatz}\\
&\bra{\psi_n^\perp} \equiv \bra{\psi^\perp(u)} = ( \ell{4}(u),\, \ell{1}(u)) \label{eq:QubitAnsatzPerp}
\end{align} 
Inserting the ansatz (\ref{eq:QubitAnsatz}), (\ref{eq:QubitAnsatzPerp}) into the divergence condition (\ref{eq:div}), the coefficients $a_n, a_{n+1},d_n$ are given by

\begin{align}
& 2(a_n - a_{n+1}) = J_- \left[ \frac{\ell{4}(u)  \ell{4}(u+\eta)}{  \ell{1}(u)  \ell{1}(u+\eta) } - \frac{\ell{1}(u)  \ell{1}(u+\eta)}{  \ell{4}(u)  \ell{4}(u+\eta) }
     \right], \nonumber \\
& 2(a_n + a_{n+1}) = J_+ \left[\frac{\ell{4}(u)  \ell{1}(u+\eta)}{  \ell{1}(u)  \ell{4}(u+\eta)} -    \frac{\ell{1}(u)  \ell{4}(u+\eta)}{  \ell{4}(u)  \ell{1}(u+\eta) }   
  \right], \label{aad}\\
& 2 d_n =  J_- \left[\frac{\ell{4}(u)  \ell{4}(u+\eta)}{  \ell{1}(u)  \ell{1}(u+\eta) } + \frac{\ell{1}(u)  \ell{1}(u+\eta)}{  \ell{4}(u)  \ell{4}(u+\eta) }
     \right], \nonumber
\end{align}
while the 
consistency condition (\ref{eq:consistency}) becomes
\begin{align}
& 4 J_z - J_+  
\left[\frac{\ell{4}(u)  \ell{1}(u+\eta)}{  \ell{1}(u)  \ell{4}(u+\eta)} +   \frac{\ell{1}(u)  \ell{4}(u+\eta)}{  \ell{4}(u)  \ell{1}(u+\eta) }   
  \right] +J_-  
 \left[\frac{\ell{4}(u)  \ell{4}(u+\eta)}{  \ell{1}(u)  \ell{1}(u+\eta) } + \frac{\ell{1}(u)  \ell{1}(u+\eta)}{  \ell{4}(u)  \ell{4}(u+\eta) } \right] =0.\label{consistency}
\end{align}
The above consistency condition can be proved,
while Eq. (\ref{aad}) can be simplified using elliptic theta function identities. 

For the convenience of calculation, we only use the notations $\ell{1}$ and $\bell{1}$ in the proof.  Using Landen's type of transformation (\ref{Landen_2}), we rewrite the anisotropy constants  
\begin{align}
J_x=\frac{\theta_1(\frac{\tau}{2}-\eta)\theta_1(\frac{\tau}{2}+\eta)}{\theta_1^2(\frac{\tau}{2})},\quad
J_y=\frac{\theta_1(\frac{1+\tau}{2}-\eta)\theta_1(\frac{1+\tau}{2}+\eta)}{\theta_1^2(\frac{1+\tau}{2})},\quad 
J_z=e^{i\pi\eta}\frac{\theta_1(\eta+\frac{1}{2}+\tau)\theta_1(\eta+\frac{1}{2})}{\theta_1(\frac{1}{2}+\tau)\theta_1(\frac{1}{2})}.
\end{align}
Using Eqs. (\ref{Riemann}) and (\ref{Double_angle_2}), we get
\begin{align}
J_-=-2e^{-i\pi\tau}\frac{\theta_1^2(\eta)}{\theta_1^2(\tau)},
\qquad J_+=-2\frac{\theta_1(\eta+\tau)\theta_1(\eta-\tau)}{\theta_1^2(\tau)}.
\end{align}
Then, we have 
\begin{align}
&- J_+  
\left[\frac{\ell{4}(u)  \ell{1}(u+\eta)}{  \ell{1}(u)  \ell{4}(u+\eta)} +   \frac{\ell{1}(u)  \ell{4}(u+\eta)}{  \ell{4}(u)  \ell{1}(u+\eta) }   
\right] +J_-  
\left[\frac{\ell{4}(u)  \ell{4}(u+\eta)}{  \ell{1}(u)  \ell{1}(u+\eta) } + \frac{\ell{1}(u)  \ell{1}(u+\eta)}{  \ell{4}(u)  \ell{4}(u+\eta) } \right]\no\\
&=2\frac{\ell{1}(\eta+\tau)\ell{1}(\eta-\tau)}{\ell{1}^2(\tau)}\left[e^{-i\pi\eta}\frac{\ell{1}(u+\tau)\ell{1}(u+\eta)}{\ell{1}(u)\ell{1}(u+\eta+\tau)}+e^{i\pi\eta}\frac{\ell{1}(u)\ell{1}(u+\eta+\tau)}{\ell{1}(u+\tau)\ell{1}(u+\eta)}\right]\no\\
&\quad -2\frac{\ell{1}^2(\eta)}{\ell{1}^2(\tau)}\left[e^{i\pi\eta}\frac{\ell{1}(u-\tau)\ell{1}(u+\eta+\tau)}{\ell{1}(u)\ell{1}(u+\eta)}+e^{-i\pi\eta}\frac{\ell{1}(u+2\tau)\ell{1}(u+\eta)}{\ell{1}(u+\tau)\ell{1}(u+\eta+\tau)}\right]\no\\
&\overset{(\ref{Riemann})}{=}-\frac{2e^{i\pi\eta}}{\ell{1}(u)\ell{1}(u+\tau)}\left[\ell{1}(u-\eta)\ell{1}(u+\eta+\tau)+e^{-2i\pi\eta}\ell{1}(u+\eta)\ell{1}(u-\eta+\tau)\right]\no\\
&\overset{(\ref{identity_1})}{=}2e^{i\pi\eta}\frac{\ell{1}(\tau)\ell{1}(2\eta)}{\ell{1}(\eta-\tau)\ell{1}(\eta)}\overset{(\ref{Double_angle_2})}{=}-4e^{i\pi\eta}\frac{\ell{1}(\eta+\frac12+\tau)\ell{1}(\eta+\frac12)}{\ell{1}(\frac12+\tau)\ell{1}(\frac12)}=-4J_z.	
\end{align}
So, the consistency condition (\ref{consistency}) is proved. From Eqs. (\ref{aad})-(\ref{consistency}), we get the expression of $a_n$
\begin{align}
	a_n&\equiv a(u)=-J_z+\frac{J_+}{2}\frac{\ell{4}(u)\ell{1}(u+\eta)}{\ell{1}(u)\ell{4}(u+\eta)}-\frac{J_-}{2}\frac{\ell{1}(u)\ell{1}(u+\eta)}{\ell{4}(u)\ell{4}(u+\eta)}\no\\
	&\overset{(\ref{Riemann})}{=}-e^{-i\pi\eta}\frac{\ell{1}(\eta+\frac{1}{2}-\tau)\ell{1}(\eta+\frac{1}{2})}{\ell{1}(\frac{1}{2}+\tau)\ell{1}(\frac{1}{2})}+{e^{-i\pi\eta}}\frac{\ell{1}(u+\eta)\ell{1}(u-\eta+\tau)}{\ell{1}(u)\ell{1}(u+\tau)}\no\\
	&\overset{(\ref{Riemann})}{=}\frac{e^{i\pi\eta}\ell{1}(\eta+\tau)\ell{1}(\eta)\ell{1}(u+\frac{1}{2})\ell{1}(u+\frac12+\tau)}{\ell{1}(\frac{1}{2}+\tau)\ell{1}(\frac{1}{2})\ell{1}(u)\ell{1}(u+\tau)}\no\\
	&\overset{(\ref{Landen_2})}{=}\frac{\bell{1}(\eta)\,\bell{2}(u)}{\bell{2}(0)\,\bell{1}(u)}.\label{res:a(u)}
\end{align}
Analogously, $a_{n+1}$ can be obtained 
\begin{align}
	a_{n+1}=a(u+\eta).
\end{align}

We can simplify the expression of $d_n$ as follows
\begin{align}
	d_n&\equiv d(u,\eta)= -J_z + \frac{J_+}{2}  
	\left[\frac{\ell{4}(u)  \ell{1}(u+\eta)}{  \ell{1}(u)  \ell{4}(u+\eta)} +   \frac{\ell{1}(u)  \ell{4}(u+\eta)}{  \ell{4}(u)  \ell{1}(u+\eta) }   
	\right]\no\\
	&\overset{(\ref{zeta;sigma;1})}{=}\frac{\bell{1}(\eta)}{\bell{1}'(0)}[\bar\zeta(\eta)-2\zeta(\eta)]+\frac{\ell{1}^2(\eta+\tau)}{\ell{1}^2(\tau)}\left[e^{i\pi\eta}\frac{\ell{1}(u+\tau)\ell{1}(u+\eta)}{\ell{1}(u)\ell{1}(u+\eta+\tau)}+e^{3i\pi\eta}\frac{\ell{1}(u)\ell{1}(u+\eta+\tau)}{\ell{1}(u+\tau)\ell{1}(u+\eta)}\right]\no\\
	&\overset{(\ref{zeta;sigma;5})}{=}\frac{\bell{1}(\eta)}{\bell{1}'(0)}[\bar\zeta(\eta)+\zeta(u)+\zeta(u+\tau)-\zeta(u+\eta+\tau)-\zeta(u+\eta)]\no\\
	&\overset{(\ref{zeta;3})}{=}\frac{\bell{1}(\eta)}{\bell{1}'(0)}[\bar\zeta(\eta)+\bar\zeta(u)-\bar\zeta(u+\eta)].\label{res:d(u)}
\end{align}


To sum up, the divergence condition (\ref{eq:div}) has the form 
\begin{align}
& h \ket{\psi(u)}  \otimes\ket{\psi(u+\eta)} = [a(u) \sigma^z\otimes \one_2  - a(u+\eta)  \one_2 \otimes \sigma^z + d(u,\eta) \one_4\,] \ket{\psi(u)} \otimes \ket{\psi (u+\eta)}
\label{eq:divplus}
\end{align}
where $a(u),d(u,\eta)$ are given by (\ref{res:a(u)}), (\ref{res:d(u)}).

Note now that the oddness property of the $\bell{1}(-\eta)= - \bell{1}(\eta) $ implies an equivalent form of the divergence condition for the 
``downhill" gradient, 
\begin{align}
& h \ket{\psi(u)} \otimes\ket{\psi(u-\eta)} = [-a(u) \sigma^z \otimes  \one_2  + a(u-\eta)  \one_2 \otimes \sigma^z  + d(u,-\eta) \one_4] \ \ket{\psi(u)} \otimes \ket{\psi (u-\eta)}
\label{eq:divminus}
\end{align}

From the definition (\ref{eq:QubitAnsatzPerp}) and properties of the elliptic functions in Eqs. (\ref{FourTheta}), (\ref{periodicity}),
we obtain
\begin{align}
\bra{\psi_n^\perp}& = ( \ell{4}(u),\, \ell{1}(u))= -i  e^{i \pi (u+\frac{\tau}{2})}   ( \ell{1}(u+\tau),\, \ell{4}(u+\tau))\no\\
&= i  e^{i \pi (u+\frac{\tau}{2})} ( \ell{1}(u+\tau+1),\, -\ell{4}(u+\tau+1)).
\end{align} 
With the help of the properties of $\bell{\alpha}(u)$ and $\bar{\zeta}(u)$, we note that  
\begin{align}
& a(u+\tau+1) = - a(u),\\
& a(u+\eta+\tau+1) = - a(u+\eta),\\
& d(u+\tau+1,\eta) = d (u,\eta),
\end{align}
Consequently, due to $h^T=h$, the transposed form of (\ref{eq:divplus}) implies yet another divergence relation for bra vectors
\begin{align}
& \bra{\psi_n^\perp} \otimes \bra{\psi_{n+1}^\perp}  h =
\bra{\psi_n^\perp} \otimes \bra{\psi_{n+1}^\perp}  
 [-a(u) \si^z\otimes\one_2 + a(u+\eta) \one_2\otimes\si^z + d(u,\eta)\one_4].
\label{eq:div-perp}
\end{align}

\section{Divergence condition for $XXZ$ chain}
\label{S-III}

Divergence condition for $XXZ$ model, analogous  to (\ref{eq:divcond1}) for $XY\!Z$ model, was already given elsewhere 
\cite{2016PRAconCarlo,2017PopkovSchutzHelix}. Here we show how to  derive 
it directly from the  (\ref{eq:divcond1}) in the limit $XY\!Z$ $\rightarrow$ $XXZ$. 
Note that only when $\tau$ is purely imaginary and $\eta$ is real or purely imaginary, the $XY\!Z$ bulk Hamiltonian is Hermitian, specifically as follows 
\begin{itemize}
	\item when $\rm{Im}[\eta]=\rm{Re}[\tau]=0$,\,\, $J_x\geq J_y\geq J_z$, 
	\item when $\rm{Re}[\eta]=\rm{Re}[\tau]=0$,\,\, $J_x\leq J_y\leq J_z$.
\end{itemize}
When $\tau\to+i \infty$, the system degenerates into $XXZ$ chain as follows:
\begin{align}
	&J_x\to1,\quad J_y\to1,\quad J_z\to\cos(\pi\eta)\leq 1,\quad \mathrm{Im}[\eta]=0,\\
	&J_x\to 1,\quad J_y\to1,\quad J_z\to\cos(\pi\eta)\geq 1,\quad \mathrm{Re}[\eta]=0.
\end{align}

Assume that $u=\frac{\tau}{2}+\tilde u$. In the limit $\tau\to+i  \infty$, $u\to {\tilde u} +  i \infty$, we have
\begin{align}
	&\lim_{\tau\to +i\infty}e^{\frac{i\pi\tau}{4}}\bell{1}(u)=\lim_{\tau\to +i\infty}\ell{1}(u)
	=ie^{-i\pi\tilde u},\no\\
	&\lim_{\tau\to +i\infty}e^{\frac{i\pi\tau}{4}}\bell{2}(u)=e^{-i\pi\tilde u},\quad\lim_{\tau\to +i\infty}\ell{4}(u)=1, \quad \tilde u\,\,\mbox{is finite}.
\end{align}
Under such assumption, we get 
\begin{align}
	&\lim_{\tau\to +i\infty}a(u)=-i\sin(\pi\eta),\\_{\rm l}
	&\lim_{\tau\to +i\infty}d(u,\eta)=\cos(\pi\eta),\\
	&\lim_{\tau\to +i\infty}-\frac{\ell{4}(u)}{\ell{1}(u)}=e^{i\pi(\tilde u+\frac{1}{2})}.
\end{align} 
Now we see the spin-helix structure and the divergence condition (\ref{eq:divplus}) degenerates into the $XXZ$ type \cite{2021PhantomLetter,2021PhantomBetheAnsatz,2021PhantomLong}
\begin{align}
&h^{XXZ} \ket{\tilde\psi(\tilde u)}  \otimes\ket{\tilde\psi(\tilde u+\eta)} = [-i\sin(\pi\eta) \sigma^z\otimes \one_2  + i\sin(\pi\eta)  \one_2 \otimes \sigma^z + \cos(\pi\eta) \one_4\,] \ket{\tilde\psi(\tilde u)} \otimes \ket{\tilde\psi (\tilde u+\eta)},
\end{align}
where $h^{XXZ}=\sigma^x\otimes \sigma^x+\sigma^y\otimes \sigma^y+\cos(\pi\eta)\,
\sigma^z\otimes \sigma^z$ and $\ket{\tilde\psi(u)}=(1,\,e^{i\pi( u+\frac{1}{2})})^T$.

\section{Obtaing the left auxiliary vector $\bra{w_{\rm l}}$. Proving 
Eq. (\ref{res:wl}). }
\label{S-IV}

Our Ansatz for Zeno-limit  MPA of a dissipative spin chain of length $N+2$, with sites numbering  $0,1,\ldots ,N+1$ has the usual form
\begin{align}
&\rho_{NESS}= \rho_{\rm l} \otimes R_{NESS} \otimes \rho_{\rm r}\\
&R_{NESS}= \frac{\Omega_N \Omega_N^\dagger}{{\rm Tr} (\Omega_N \Omega_N^\dagger)}\\
&\Omega_N=\bra{w_{\rm l}} \mathbf{L}_1 \mathbf{L}_2 \ldots \mathbf{L}_N \ket{w_{\rm r}}
\end{align}
where $ \rho_{\rm l} = \sum\limits_{\al} n_{\rm l}^\al \si^\al$, $ \rho_{\rm r} = \sum\limits_{\al} n_{\rm r}^\al \si^\al$
are targeted left and right boundary magnetizations, $|\vec{n}_{\rm l,r}|=1$,  $\mathbf{L}_n=\sum_\alpha L^\alpha_n \sigma^\alpha_n$ are site-dependent Lax matrices
satisfying the divergence relation
\begin{align}
&[h_{n,n+1},\mathbf{L}_n \mathbf{L}_{n+1}] = 2i(I \mathbf{L}_{n+1}-\mathbf{L}_n I).
\label{eq:bulk}
\end{align}
and $\bra{w_{\rm l}}$,  $\ket{w_{\rm r}}$ are suitable vectors in auxiliary space guaranteeing the commutation of the 
$R_{NESS}$ with the dissipation-projected $XY\!Z$ Heisenberg Hamiltonian
\begin{align}
&[R_{NESS},h_D]=0, \label{comm-hD}\\
&h_D= \sum_{n=1}^{N-1} h_{n,n+1} + \sum_{\al=x,y,z} J_\al n_{\rm l}^\al \si^\al_1 +  \sum_{\al=x,y,z} J_\al n_{\rm r}^\al \si^\al_N.
\label{hD}
\end{align}
Nonzero matrix elements of the Lax matrix $L_n$ is given by 
\begin{align}
& (L_n)_{j,j} = (L_n)_{j,j+1} = \frac{1}{\ka(u_{n-2(j-1)})} \ket{\psi(u_{n-2(j-1)})} \bra{\psi^\perp(u_{n-2(j-1)})},\\
&u_{n} = u  + n, \eta \label{def:uk} \\
&\ka(u) = - i a(u) \ell{1}(u) \ell{4}(u),
\end{align}
where $\bra{\psi^\perp(u)}$,  $\ket{\psi(u)}$ are given by   (\ref{eq:QubitAnsatzPerp}), (\ref{eq:QubitAnsatz}), and $u$ is some initial phase to be determined later, see (\ref{z0-1}), (\ref{z0-2}). 
Explicitly,  the  components $L_n^\al$ and the matrix $I$ are given by 
\begin{align}
& L^\alpha_n = \sum_{j=1}^{N+1} s^\alpha_{n-2(j-1)} \ket{j}\bra{j}+\sum_{j=1}^{N} s^\alpha_{n-2(j-1)} \ket{j}\bra{j+1}, \label{res:LnForm}\\
& s^z (u) = \frac{i}{a(u)},\\
& s^x (u) - i  s^y (u)  = \frac{i}{a(u)}  \frac{\ell{1}(u)}{\ell{4}(u)},\\
& s^x (u) + i  s^y (u)  = -\ \frac{i}{a(u)}  \frac{\ell{4}(u)}{\ell{1}(u)},\\
& I = \sum_{j=1}^{N+1} \ket{j}\bra{j}-\sum_{j=1}^{N} \ket{j}\bra{j+1}.\, 
\end{align}
Denoting $J_\al n_{\rm l/r}^\al=h_{\rm l/r}^\al$, and following the procedure outlined in 
\cite{PPZ-PRL,PPZ-PRE}, from (\ref{comm-hD})
we get  two conditions to be satisfied, 
\begin{align}
& \bra{w_{\rm l}}  V_{\rm l}  L_{2} \ldots L_{N}\ket{w_{\rm r}}=0 \label{condR-mildOper0}\\
& \bra{w_{\rm l}}  L_{1} \ldots L_{N-1} V_{\rm r} \ket{w_{\rm r}}=0,  \label{condR-mildOper}
\end{align}
where
\begin{align}
& V_{\rm l} =  \vec{h}_{\rm l}\cdot \vec{L}_1 +  i \ I,\\
& V_{\rm r} =  \vec{h}_{\rm r}\cdot\vec{L}_N -  i \ I.
\end{align}
The condition (\ref{condR-mildOper0}) can be  satisfied in its local form
\begin{align}
	& \bra{w_{\rm l}}  V_{\rm l} = 0
\end{align}
with an appropriate  choice of  $u_1$  in (\ref{def:uk}) and $ \bra{w_{\rm l}}$.  
Namely, let
constants $\al_k^-$ parametrize the left boundary field of the 
integrable $XY\!Z$ Hamiltonian in the standard way. It turns out that boundary field of the form 
$ (\vec{h}_{\rm l})_\al= J_\al   (\vec{n}_{\rm l})_\al$ where $(\vec{n}_{\rm l})_\al$ are components of a unit vector $|\vec{n}_{\rm l}|=1$
correspond to a choice 
\begin{align}
	&\{ \al_1^{-} , \al_2^{-} ,\al_3^{-} \}= \left\{ \eta, \frac{1}{2} + i v_1,  \frac{\tau}{2} + v_2\right\},
	\label{alphaminus}
\end{align}
where $v_1,v_2$ are real numbers.
Let us choose  $u_1=u+\eta$ as 
\begin{align}
	& u_1 =  \frac{1}{2} + \al_1^{-}  + \al_2^{-} + \al_3^{-} \label{z0-1}
\end{align}
or 
\begin{align}
	& u_1  =\frac{1}{2} + \al_1^{-}  - \al_2^{-} - \al_3^{-} \label{z0-2}.
\end{align}
Then, one can show that 

\begin{align}
	& ( V_{\rm l})_{11}=  ( V_{\rm l})_{23} =0,  \label{eq:VLrelations}\\
	&( V_{\rm l})_{12}=  -( V_{\rm l})_{22}  \neq 0, 
\end{align}
and consequently  $\bra{w_{\rm l}}  V_{\rm l} = 0$ is satisfied with the choice 
\begin{align}
	& \bra{w_{\rm l}} = (1,1,0,0,\ldots ,0 ),
\end{align} 
so, remarkably, the ``left boundary"  condition  $\bra{w_{\rm l}}  V_{\rm l} = 0$
is satisfied with the universal (anisotropy independent) choice of $w_{\rm l}$.

\bigskip

\textbf{Proof of (\ref{eq:VLrelations}):}
The left boundary field of the integrable $XY\!Z$ Hamiltonian is standardly parameterized as \cite{Yang2006,Cao2013,OffDiagonal}
\begin{align}
& (\vec{h}_{\rm l})_x=e^{-i\pi(\sum_{k=1}^3\alpha_k^--\frac{\tau}{2})}\frac{\bell{1}(\eta)}{\bell{1}(\frac\tau2)}\prod_{k=1}^3\frac{\bell{1}(\alpha_k^--\frac{\tau}{2})}{\bell{1}(\alpha_k^-)},\no\\
&(\vec{h}_{\rm l})_y = e^{-i\pi(\sum_{k=1}^3\alpha_k^--\frac{1+\tau}{2})}\frac{\bell{1}(\eta)}{\bell{1}(\frac{1+\tau}{2})}\prod_{k=1}^3\frac{\bell{1}(\alpha_k^--\frac{1+\tau}{2})}{\bell{1}(\alpha_k^-)},\no\\
&(\vec{h}_{\rm l})_z= \frac{\bell{1}(\eta)}{\bell{1}(\frac12)}\prod_{k=1}^3\frac{\bell{1}(\alpha_k^--\frac{1}{2})}{\bell{1}(\alpha_k^-)},\label{boundary;fields}
\end{align}
and Eqs. (\ref{eq:VLrelations}) can then be written as 
\begin{align}
& F(u_1) = -  a(u_1), \label{eq1}\\
& F(u_1-2 \eta) =  a(u_1-2 \eta)\,,  \label{eq2}
\end{align}
where
\begin{align}
	F(u)=(\vec{h}_{\rm l})_z+\frac{(\vec{h}_{\rm l})_x}{2}\left[-\frac{\ell{4}(u)}{\ell{1}(u)}+\frac{\ell{1}(u)}{\ell{4}(u)}\right]+\frac{i(\vec{h}_{\rm l})_y}{2}\left[\frac{\ell{4}(u)}{\ell{1}(u)}+\frac{\ell{1}(u)}{\ell{4}(u)}\right].\label{def:F(u)}
\end{align}
Using Eqs. (\ref{Riemann}), (\ref{Landen_1}) and (\ref{Double_angle_1}), we get 
\begin{align}
&-\frac{\ell{4}(u)}{\ell{1}(u)}+\frac{\ell{1}(u)}{\ell{4}(u)}=2e^{i\pi(u+\frac{1+\tau}{2})}\frac{\bell{1}(u+\frac{1+\tau}{2})\bell{1}(\frac{\tau}{2})}{\bell{1}(u)\bell{1}(\frac12)},\no\\
&-\frac{\ell{4}(u)}{\ell{1}(u)}-\frac{\ell{1}(u)}{\ell{4}(u)}=2e^{i\pi(u+\frac{1+\tau}{2})}\frac{\bell{1}(u+\frac{\tau}{2})\bell{1}(\frac{1+\tau}{2})}{\bell{1}(u)\bell{1}(\frac12)}.\label{zeta}
\end{align}
With the help of Eqs. (\ref{Riemann}) and (\ref{zeta}), we finally simplify the expression of $F(u)$ after tedious calculations 
\begin{align}
F(u)
&=a(u)+2\frac{\bell{1}(\eta)\prod_{l=0}^3\bell{1}\left(\tfrac{2u-2\chi_{\rm l}-1}{4}\right)}{\bell{1}(\frac12)\bell{1}(u)\prod_{k=1}^3\bell{1}(\alpha_k^-)}\no\\
&=-a(u)+2\frac{\bell{1}(\eta)\prod_{l=0}^3\bell{1}\left(\tfrac{2u+2\chi_{\rm l}-1}{4}\right)}{\bell{1}(\frac12)\bell{1}(u)\prod_{k=1}^3\bell{1}(\alpha_k^-)},\label{F;expression}
\end{align}
where 
\begin{align}
&\chi_0=-\alpha_1^--\alpha_2^--\alpha_3^-,\quad \chi_1=\alpha_2^-+\alpha_3^--\alpha_1,\no\\ &\chi_2=\alpha_1^--\alpha_2^-+\alpha_3^-, \quad
\chi_3=\alpha_1^-+\alpha_2^--\alpha_3^-.\label{chi}
\end{align}
The value of $u_1$ is 
\begin{align}
&u_1=\frac12-\chi_0=\frac12+\chi_1+2\eta,\no\\ \mbox{or}\,\,\, &u_1=\frac12-\chi_1=\frac12+\chi_0+2\eta.
\end{align}
From the explicit expression of $F(u)$ in (\ref{F;expression}), it is straightforward that $$F(u_1)=-a(u_1),\quad F(u_1-2\eta)=a(u_1-2\eta).$$ 
Eqs. (\ref{eq1})--(\ref{eq2}) are thus proven.

Finally, parameter $u$ from the Eqs.(3),(11) in the main text can be found as
 $u= u_1 - \eta$. Accounting for (\ref{alphaminus}),  we get 
\begin{align}
&u=\frac{\tau}{2}+ i v_1 + 1 + v_2 = x_{\rm l} +  i y_{\rm l}, \\
&y_{\rm l} = v_1 + \frac{\tau}{2 i}, \quad x_{\rm l} = 1+ v_2. \label{Def:xlyl}
\end{align}
The components of the targeted left boundary magnetization  are related to components of  the boundary fields via 
$n_{\rm l}^\al=h_{\rm l}^\al/J_\al $. 
Substituting this into (\ref{boundary;fields}), using (\ref{Def:anisotropy}),  (\ref{Def:xlyl})
and using basis relations for elliptic theta-functions, 
 one obtains 
\begin{align}
&n_{\mu}^x = -\frac{ \bell{2}(i y_{\mu})}{\bell{3}(i y_{\mu})} \   \frac{ \bell{1}(x_{\mu})}{\bell{4}(x_{\mu})}, \nonumber\\
&n_{\mu}^y = - i \frac{ \bell{1}(i y_{\mu})}{\bell{3}(i y_{\mu})} \   \frac{ \bell{2}(x_{\mu})}{\bell{4}(x_{\mu})},  \nonumber\\  
&n_{\mu}^z =- \frac{ \bell{4}(i y_{\mu})}{\bell{3}(i y_{\mu})} \   \frac{ \bell{3}(x_{\mu})}{\bell{4}(x_{\mu})}\,. \nonumber
\end{align}
i.e. the Eq.~(\ref{eq:nvec-parametrization}) for $\vec{n}_{\rm l}$
and Eq.~(\ref{res:CommSolution}) in the main text. 

We can  check that  $n_{\mu}^x,n_{\mu}^y,n_{\mu}^z$ are indeed 
components of a unit vector.
We shall use an identity (\cite{WatsonBook}, p. 487):
\begin{align}
\bell{3}(x+y)\bell{3}(x-y)\bell{4}^2(0)=\bell{3}^2(x)\bell{4}^2(y)-\bell{2}^2(x)\bell{1}^2(y).\label{addition;formula}
\end{align}
For purely imaginary $\tau$ and  real $x$, $\bell{\alpha}(x),\,\alpha=1,2,3,4$ and $\bell{\beta}(ix),\,\beta=2,3,4$ are real,
while $\bell{1}(ix)$ is purely imaginary. Using Eq. (\ref{addition;formula}), we get 
\begin{align}
&(n_{\mu}^{x})^2+(n_{\mu}^{y})^2+(n_{\mu}^{z})^2-1\nonumber\\
&= \frac{1}{\bell{3}^2(i y_{\mu})\bell{4}^2(x_{\mu})}\left[\bell{4}^2(i y_{\mu})\bell{3}^2(x_{\mu})-\bell{1}^2(i y_{\mu})\bell{2}^2(x_{\mu})+\bell{2}^2(i y_{\mu})\bell{1}^2(x_{\mu})-\bell{3}^2(i y_{\mu})\bell{4}^2(x_{\mu})\right]\nonumber\\
&=\frac{\bell{4}^2(0)\bell{3}(x_\mu+iy_\mu)}{\bell{3}^2(i y_{\mu})\bell{4}^2(x_{\mu})}\left[\bell{3}(x-iy)-\bell{3}(iy_\mu-x_\mu)\right]\nonumber\\
&=0.
\end{align}

To obtain the part of the Eq.(\ref{eq:nvec-parametrization}) in the main text concerning $\vec{n}_{\rm r}$ we 
 choose 

\begin{align}
	&\{ \al_1^{+} , \al_2^{+} ,\al_3^{+} \}= \left\{ \eta, \frac{1+\tau}{2} - i y_{\rm r},  \frac{\tau}{2} + x_{\rm r}\right\},
\end{align}
and proceed analogously.

\bigskip

\section{Calculating the right auxiliary vector $\ket{w_{\rm r}}$. Showing consistency of the recurrence (\ref{recurrence}).}
\label{S-V}

While the universal form of  $\bra{w_{\rm l}}$ satisfying $\bra{w_{\rm l}} V_{\rm l} =0$ can be found, 
the analogical local right boundary condition  $V_{\rm r} \ket{w_{\rm r}}=0$ cannot be satisfied, since, in general, 
the operator $V_{\rm r}$ does not have zero eigenvalues. 
Instead, a milder condition 

\begin{align}
	& \bra{w_{\rm l}} (L_1)_{\al_1} (L_2)_{\al_2} \ldots (L_{N-1})_{\al_{N-1}} V_{\rm r} \ket{w_{\rm r}}=0,  \label{condR-mildExpanded}
\end{align}
valid for any $\al_1, \ldots \al_{N-1}\in\{x,y,z\}$, 
can be satisfied by a nontrivial choice

\begin{align}
	&V_{\rm r} \ket{w_{\rm r}} =\ket{1,-1,1,-1,\ldots } \label{res-wR-mild}.
\end{align}
Indeed, due to a special band form of all $L_n^\al$  with equal elements within each row (\ref{res:LnForm})
we have 
\begin{align*}
	& \bra{w_{\rm l}} (L_1)_{\al_1} (L_2)_{\al_2} \ldots (L_{N-1})_{\al_{N-1}} V_{\rm r} \ket{w_{\rm r}}=\\
	& \bra{w_{\rm l}} (L_1)_{\al_1} (L_2)_{\al_2} \ldots (L_{N-1})_{\al_{N-1}} \ket{1,-1,1,-1,...}\\
	& =\bra{w_{\rm l}}(L_1)_{\al_1} (L_2)_{\al_2} \ldots (L_{N-2})_{\al_{N-2}} \ket{0, \ldots ,0,*}\\
	& =\braket{1,1,0,\ldots|0,0,*,*,  \ldots ,* }=0  
\end{align*}
where we mark by the asterisk sign $*$ potentially nonzero entries of the right ket vector. 
If $\det V_{\rm r} \neq 0$ (the generic case), a unique nontrivial 
solution for $\ket{w_{\rm r}}$ exists, given by the explicit recurrence (\ref{recurrence}) in the main text. 

On the other hand, under  special conditions rendering  $\det[ V_{\rm r}] = 0$, which entails 
$(V_{\rm r})_{nn}=0$ for some $n$, 
the recurrence (\ref{recurrence}) in the main text breaks down. Instead, the vector $\ket{w_{\rm r}}$ can be chosen 
such as to satisfy the relation (\ref{condR-mildOper})  locally, as shown in the next section.

\section{ Spin-helix states analogues for $XY\!Z$ model.  The case of $\det V_{\rm r}=0$.}
\label{S-VI}

Elements of $ V_{\rm r}$ in the $k$-th row have the form 
\begin{align}
	&\left(V_{\rm r}  \right)_{kk} = \text{const} \times  \left[\frac{G(v )}{a(v)} + 1 \right],\\
	&\left(V_{\rm r}  \right)_{k-1,k} = \text{const} \times  \left[ -\frac{G(v+2\eta )}{a(v+2\eta)} + 1 \right],\\
	&v=u_1 + (N+1-2k)\eta, \label{res:PhantomConditionXYZ} \\
	&k=1,2, \ldots N+1
\end{align}
where $G(x)$ is obtained from $F(x)$ in (\ref{def:F(u)}) by substitutions $\al_k^{-} \rightarrow \al_k^{+}$.

An inspection of the above shows that both elements $\left(V_{\rm r}  \right)_{kk} =\left(V_{\rm r}  \right)_{k-1,k} =0 $
vanish if one of the two conditions is satisfied,
\begin{align}
	&v =\frac{1}{2}  -    \al_1^{+} -    \al_2^{+}  +   \al_3^{+} \label{res-v},\\
	&v =\frac{1}{2}  -    \al_1^{+} +    \al_2^{+}  -   \al_3^{+} \label{res-v-1},\\
	&\text{where}\quad \{ \al_1^{+} , \al_2^{+} ,\al_3^{+} \}= \left\{ \eta, \frac{1+\tau}{2} - i y_{\rm r},  \frac{\tau}{2} + x_{\rm r}\right\}.
	\label{alphaplus}
\end{align}
By doing some simple substitutions on $F(u)$ and using the expression in (\ref{F;expression}), the above property can be easily proved. The 
condition (\ref{res:PhantomConditionXYZ}) turns out  to be equivalent to 
\begin{align}
	& x_{\rm r}= x_{\rm l} +(N+1-2 M_0)\eta\,, \quad y_{\rm r}= y_{\rm l}\,, \qquad M_0 = 0,1,\ldots N,\label{res-uR-Phantom}
\end{align}
used to get the factorized periodically modulated state (\ref{res:ell}) in the main text, for $M_0=0$.
Then the equation (\ref{res-wR-mild}) cannot be solved because of singularity in (\ref{recurrence}),
while  the equation 
\begin{align}
	&V_{\rm r} \ket{w_{\rm r}} = V_{\rm r} (r_1,r_2, \ldots r_{N+1})^T =0  \label{res-wR-local}
\end{align}
 is readily solved with
\begin{align}
	& r_k= \de_{k,M_0+1}.  \label{res-wR-Phantom}
\end{align}

Now, the periods of the $G(u)$ are 
$G(u+2 n_0 + 2 \tau m_0)= G(u)$ where $n_0,m_0$ are integers. It follows that 
if periodicity conditions are satisfied 
\begin{align}
	& \eta M_1 =2 n_0 + 2 \tau m_0\,,\\
	&M_1 < N\,,  \label{cond:Multiple-wR}
\end{align}
there could be several rows of $V_{\rm r}$ which become zero simultaneously. Then, the zero eigenvalue of  $V_{\rm r}$ has degeneracy,
and a unique solution of (\ref{res-wR-local}) for $\ket{w_{\rm r}}$ does not exist.  Proper consideration of this case needs considering subleading corrections to the Zeno limit of NESS and is out of our present scope.

\section{Derivation of Eq. (\ref{res:ell}). Obtaining one-point observables (magnetization profile)  in Fig. 1. }
\label{S-VII}

Pure NESS Eq. (\ref{SHSell}) in the main text has the form of  a factorized product of the qubits Eq.(\ref{eq:spinor}). 
Let  us calculate the expectations 
$\langle \si_n^z \rangle $, $\langle \si_n^+ \rangle $.
Using (\ref{eq:spinor}) and denoting $u_n=u +n \eta$, we obtain
\begin{align}
	& \langle \si_n^+ \rangle
	=-\frac{\ell{1}^*(u_n)\ell{4}(u_n)}{\ell{1}^*(u_n)\ell{1}(u_n)+\ell{4}^*(u_n)\ell{4}(u_n)},\\
& \langle \si_n^z \rangle=\frac{\ell{1} (u_n) \ell{1}^* (u_n) - \ell{4} (u_n) \ell{4}^* (u_n) }{\ell{1} (u_n) \ell{1}^* (u_n) + \ell{4} (u_n) \ell{4}^* (u_n)},
\end{align}
where $^*$ denotes complex conjugation. Using $\ell{1}^* (u)= \ell{1} (u^*)$,  $\ell{4}^* (u)= \ell{4} (u^*)$ and the formulae for elliptic 
functions (\ref{Derkachev1}-\ref{Derkachev3}),
we readily obtain
\begin{align}
	&\langle \si_n^x \rangle = 2 {\rm Re}[\langle \si_n^+ \rangle]=- \frac {\bell{1} (x) \bell{2} ( i y) }   {\bell{4} (x) \bell{3} (i y) } , \\
	&\langle \si_n^y \rangle  = 2 {\rm Im}[\langle \si_n^+ \rangle]=- i \frac {\bell{2} (x) \bell{1} ( i y) }   {\bell{4} (x) \bell{3} (i y) } , \\
&\langle \si_n^z \rangle = -  \frac {\bell{3} (x) \bell{4} ( i y) }   {\bell{4} (x) \bell{3} (i y) } , 
\end{align}
where $x  = {\rm Re}[u + n \eta ]=  {\rm Re}[u] + n \eta$,  $y  = {\rm Im}[u+ n \eta]={\rm Im}[u]$, assuming $\eta$ being real.   
Finally, using the relation to well known Jacobi elliptic functions ${\rm sn, cn,  dn}$ and denoting  
 \begin{align}
&k = \left(\frac{\bell{2} (0) }{\bell{3} (0) }\right)^2\,, \quad k'=\sqrt{1-k^2}= \left(\frac{\bell{4} (0) }{\bell{3} (0) }\right)^2, \quad K_k = \frac12 \pi \ (\bell{3} (0))^2\,,\\
&A_x= -\sqrt{k} \ \frac { \bell{2} ( i\  {\rm Im}[u] ) }   {\bell{3} (i \ {\rm Im}[u] ) }, \quad A_y = -i \sqrt{\frac{k}{k'}} \  \frac { \bell{1} ( i \ {\rm Im} [u])}{\bell{3} (i \ {\rm Im}[u] ) }\,,\quad
A_z = -\frac{1}{\sqrt{k'}}\  \frac { \bell{4} ( i\  {\rm Im}[u])} {\bell{3} (i\ {\rm Im}[u])}\,,
\end{align}
Note that in our physical case $0\leq k \leq 1$ is real.  For general case (e.g. for complex $\tau$ parameter) one can use  
\begin{align}
&A_x= -\frac{\bell{3} (0) }{\bell{2} (0) } \ \frac { \bell{2} ( i\  {\rm Im}[u] ) }   {\bell{3} (i \ {\rm Im}[u] ) }, 
\quad A_y = -i    \frac{\bell{4} (0) }{\bell{2} (0) } \  \frac { \bell{1} ( i \ {\rm Im} [u])}{\bell{3} (i \ {\rm Im}[u] ) }\,,\quad
A_z = -\frac{\bell{4} (0) }{\bell{3} (0) }\  \frac { \bell{4} ( i\  {\rm Im}[u])} {\bell{3} (i\ {\rm Im}[u])}\,,
\end{align}
valid for arbitrary choice of $\tau$ and $\eta$.
We finally obtain
\begin{align}
	&\langle \si_n^x \rangle=A_x \ {\rm sn}(2 K_k ({\rm Re}[u] + n \eta),k )\,, \\
	&\langle \si_n^y \rangle =A_y \ {\rm cn}(2 K_k ({\rm Re}[u] + n \eta),k )\,, \\
	&\langle \si_n^z \rangle =A_z \ {\rm dn}(2 K_k ({\rm Re}[u] + n \eta),k )\,, 
\end{align}
leading to Eq. (\ref{res:ell}) in the main text.




\begin{thebibliography}{99}
\bibitem{footnote} Since $V_{\rm r}$ is a triangular matrix, its determinant is given as a product of its diagonal elements.
\end{thebibliography}
\end{document}